\documentclass[%
 reprint,%
 amssymb, amsmath,%
 aip,apl,%
]{revtex4-1}
\usepackage[utf8x]{inputenc}
\usepackage{graphicx}
\usepackage{docs}%
\usepackage{bm}%
\usepackage{amsmath}
\usepackage[colorlinks=true,linkcolor=blue]{hyperref}%
\hypersetup{breaklinks=true}
\expandafter\ifx\csname package@font\endcsname\relax\else
 \expandafter\expandafter
 \expandafter\usepackage
 \expandafter\expandafter
 \expandafter{\csname package@font\endcsname}%
\fi
\newcommand{\MF}{}
\newcommand{\MFtwo}{}

\newcommand{\inv}{${}^{-1}$}
\newcommand{\invt}{${}^{-2}$}

\begin{document}

\title{ \MF{Two-well quantum cascade laser optimization by non-equilibrium Green's function modelling}}

\author{M.~Francki\'e\affiliation{}}
\email{martin.franckie@phys.ethz.ch}
\affiliation{Institute for Quantum Electronics, ETH Z\"urich, Auguste-Piccard-Hof 1, 8093
Z\"urich, Switzerland}
\author{L.~Bosco\affiliation{}}
\affiliation{Institute for Quantum Electronics, ETH Z\"urich, Auguste-Piccard-Hof 1, 8093
Z\"urich, Switzerland}
\author{M.~Beck}
\affiliation{Institute for Quantum Electronics, ETH Z\"urich, Auguste-Piccard-Hof 1, 8093
Z\"urich, Switzerland}
\author{C.~Bonzon}
\affiliation{Institute for Quantum Electronics, ETH Z\"urich, Auguste-Piccard-Hof 1, 8093
Z\"urich, Switzerland}
\author{E.~Mavrona}
\affiliation{Institute for Quantum Electronics, ETH Z\"urich, Auguste-Piccard-Hof 1, 8093
Z\"urich, Switzerland}
\author{G.~Scalari}
\affiliation{Institute for Quantum Electronics, ETH Z\"urich, Auguste-Piccard-Hof 1, 8093
Z\"urich, Switzerland}
\author{A.~Wacker}
\affiliation{Mathematical Physics and NanoLund, Lund University, Box 118, 22100
Lund, Sweden}
\author{J.~Faist}
\affiliation{Institute for Quantum Electronics, ETH Z\"urich, Auguste-Piccard-Hof 1, 8093
Z\"urich, Switzerland}

\date{15 September 2017}%
\revised{18 December 2017}%
\accepted{19 December 2017 -- Appl. Phys. Lett. \textbf{112}, 021104 (2018)}%

\begin{abstract}
We present a two-quantum well THz intersubband laser operating up to 192 K. The structure has been optimized with a non-equilibrium Green's function \MF{model. The result of this optimization was confirmed experimentally by growing, processing and measuring a number of proposed designs}. At high temperature ($T>200$ K), the simulations indicate that lasing fails due to a combination of electron-electron scattering, thermal backfilling, and, most importantly, \MF{re-absorption coming from broadened states}.
\end{abstract}

\maketitle
%
%

Terahertz quantum cascade lasers (QCLs)\cite{kohler_terahertz_2002} are interesting candidates for a wide variety of potential applications\cite{liang_recent_2017,williams_terahertz_2007}. However, to date, their operation is limited to 
$\sim$200 K\cite{fathololoumi_terahertz_2012} and the necessity of cryogenic cooling hinders a widespread use of these devices. In the last decade, significant scientific effort has been directed towards identifying the main temperature-degrading mechanisms\cite{jirauschek_limiting_2008,chassagneux_limiting_2012,li_temperature_2009,albo_investigating_2015}, as well as finding optimized QCL designs\cite{hu_resonant-phonon-assisted_2005,kumar_two-well_2009,wacker_extraction-controlled_2010,lin_significant_2011,dupont_phonon_2012,chan_tall-barrier_2013}. The degrading mechanisms include thermal backfilling\cite{williams_terahertz_2007,lindskog_injection_2013}, thermally activated LO phonon emission\cite{chassagneux_limiting_2012,li_temperature_2009,albo_investigating_2015,vitiello_quantum_2015}, increased broadening\cite{nelander_temperature_2008,khurgin_inhomogeneous_2008,matyas_role_2013}, and carrier leakage into continuum states\cite{albo_carrier_2015}. When numerically optimizing a design, it is important to take all of these effects into consideration, in order to ensure a close correspondence between the model and the real device. 
Combined with the fact that the optimization parameters are typically trade-offs for one another, the task is very complex. Here, typically simpler rate equation or density matrix models are used in order to more quickly sweep the parameter space\cite{danicic_optimization_2010,dupont_simplified_2010,bismuto_fully_2012}, while more advanced models, such as non-equilibrium Green's functions (NEGF) or Monte-Carlo, are used to validate and analyze the final designs\cite{yasuda_nonequilibrium_2009,matyas_temperature_2010,dupont_phonon_2012,lindskog_comparative_2014}. In contrast, in this work we will employ an advanced model directly at the optimization stage. Specifically, we shall use a NEGF model\cite{wacker_nonequilibrium_2013}, capable of accurately \MF{simulating} experimental devices\cite{dupont_phonon_2012,lindskog_comparative_2014,winge_simulating_2016} and \MF{including} the most general treatment of scattering, from all relevant processes.
\begin{figure}[htb]
\includegraphics[width = \linewidth]{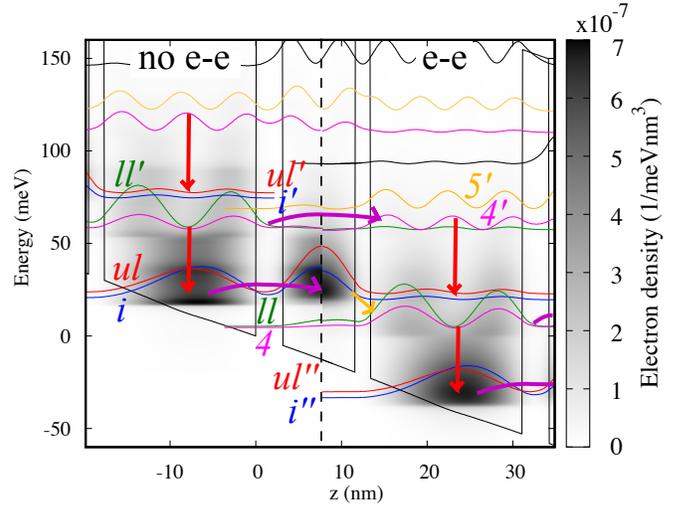}
\caption{(Color online) Band structure and Wannier-Stark states of the optimized structure with layer sequence \textbf{31}/85/\textbf{18}/\underline{177} \AA, where bold numbers indicate \MF{the injection and laser barriers.} \MF{The underline indicates the doped well, with a volume doping of 1.5$\cdot10^{17}$ cm${}^{-3}$ in the central 3 nm region, yielding an areal doping of 4.5$\cdot 10^{10}$ cm\invt}. The grayscale shows the carrier densities evaluated in the NEGF model for a lattice temperature of 200 K.
To the left (right) of the central line, the results are obtained without (with) electron-electron scattering. The red (purple) arrows indicate LO phonon (resonant tunnelling) transitions. The main laser transition is indicated by the \MF{yellow} arrow. \MF{$ul$ denotes the upper laser level in the central period, $ul'$ in the left period, and $ul''$ in the right period (and similarly for the injector state ($i$), lower laser state ($ll$), and levels $4$ and $5$).}}
\label{StatesEV2416}
\end{figure}

The goal of the optimization is to achieve the highest possible operating temperature. Thus, the gain of the active medium should be maximized at high lattice temperature, and simultaneously the external losses minimized.
The key figures for gain are inversion, oscillator strength, and line width\cite{faist_quantum_2013}. These are mainly controlled by the doping density, the energy difference $E_\text{ex}$ between the lower laser level $ll$ and the extractor state $e$, and the width of \MF{the} two barriers\MF{:} the laser and injection barriers.
Population inversion increases with doping, although a too high level promotes detrimental effects, such as electron-electron scattering. $E_\text{ex}$, which is chosen to be close to the LO phonon resonance $E_\text{LO}$ in order to have a short $ll$ lifetime, and the laser frequency $\hbar\omega$ are mainly determined by the well widths.
The laser barrier width determines the oscillator strength, which at the same time affects inversion; a more vertical transition with a larger oscillator strength, yields a lower inversion due to increased rate of non-radiative transitions from the upper laser level $ul$. These transitions broaden $ul$, and consequently also the line width.
The injection barrier limits the detrimental injection directly into $ll$, but also the injection into $ul$, and thus plays a crucial role for the population inversion.

As a starting point, we choose the shortest possible structure based on two quantum wells per period\cite{kumar_two-well_2009,scalari_broadband_2010}, in order to maximize the gain per unit length; with fewer active states per period, more carriers are expected to concentrate on the upper laser level ($ul$).
In addition, we limit the escape of carriers into continuum states by employing barriers with a high (25\%) AlAs concentration\cite{dupont_phonon_2012,chan_tall-barrier_2013,lin_significant_2011,albo_carrier_2015}. An example of a design is shown in Fig.~\ref{StatesEV2416}.
The well widths are fixed to have $E_\text{ex}\approx E_\text{LO}$ and $\hbar\omega\approx 16$ meV. The latter is chosen in order to be high enough to limit thermal backfilling, but still below the tail of the TO phonon optical absorption line. In order to limit the negative effects of impurity scattering, the 3 nm wide doping layer is placed in the central region of the widest well, where the lower laser level has its node\cite{scalari_broadband_2010}. Then, the barrier widths and the doping concentration were varied to find their optimal values for high gain at elevated temperatures. A variety of structures were evaluated both by NEGF simulations at 300 K lattice temperature and by manufacturing and characterizing experimental devices. It should be noted, that for high carrier concentrations, electron-electron (e-e) scattering will have a non-negligible impact\cite{hyldgaard_electron-electron_1996,harrison_relative_1998,manenti_monte_2003,bonno_modeling_2005,jirauschek_limiting_2008,wang_transient_2017}, and provides additional thermalization and reduction of the subband lifetimes through second order processes. This is expected to increases the current density and decreases the gain. Since we cannot fully model the e-e interactions\cite{winge_simple_2016}, we restrict the doping concentration \MF{of the grown devices to an areal doping density of 4.5$\cdot10^{10}$cm\inv~(corresponding to a volume doping density of $1.5\cdot10^{17}$ cm${}^{-3}$ of the doped 3 nm region, and an average period volume density of $\sim 1.4\cdot10^{16}$ cm${}^{-3}$)}, where we expect the effect to be moderate.

\begin{figure}
\includegraphics[width=\linewidth]{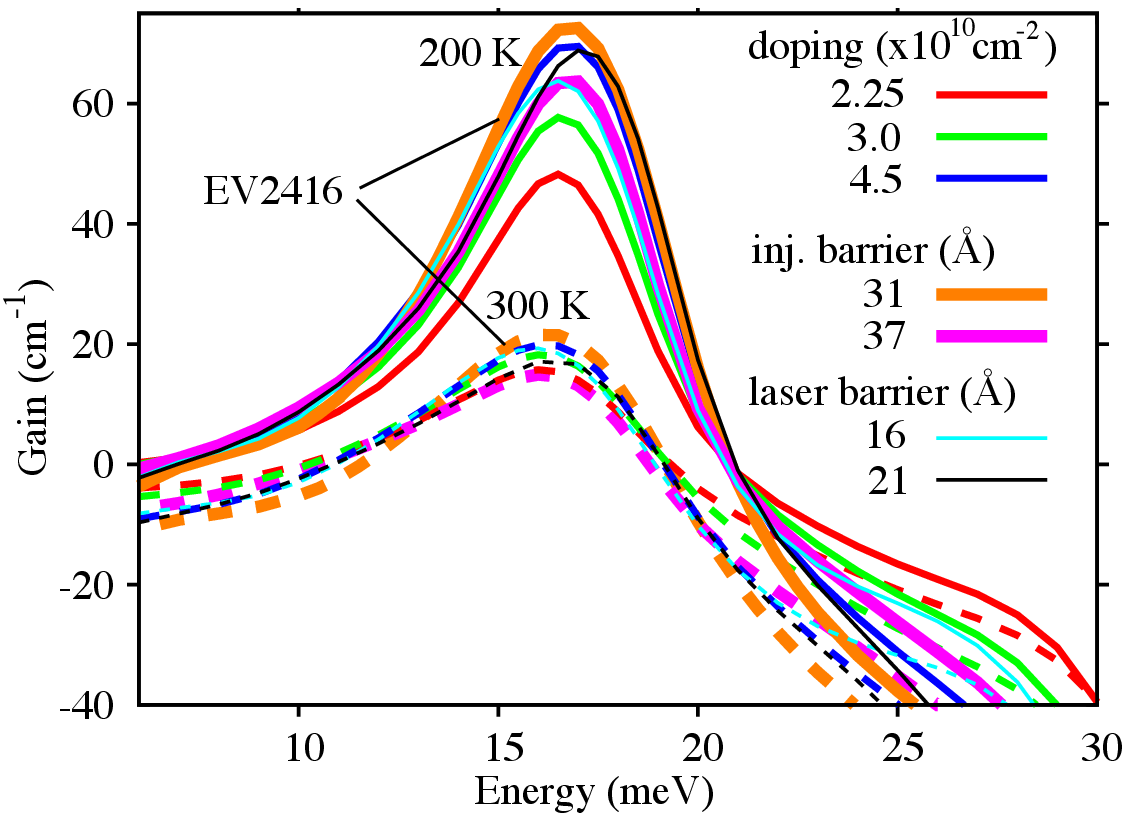}\\
\vspace*{0.2cm}
\includegraphics[width=\linewidth]{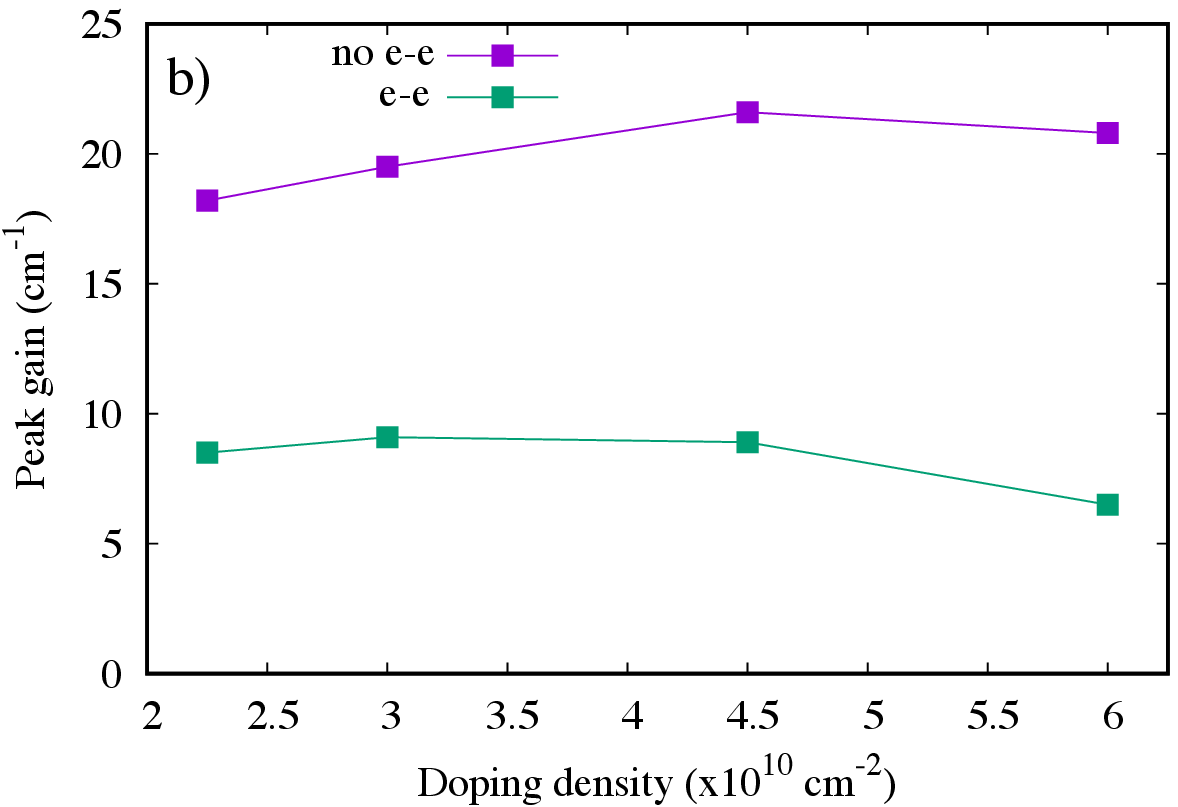}
\caption{(Color online) (a) Gain for a selected set of samples at lattice temperatures of  200 K (full lines) and 300 K (dashed lines). The top three designs have the nominal layer sequence \textbf{34}/85/\textbf{18}/\underline{177} \AA ~and varying doping densities. The following two designs have constant doping density of 4.5$\cdot 10^{10}$cm\invt, and varying injection barrier width compared to the nominal design. The bottom two designs have varying laser barrier width. The thicker lines show gain of the best design EV2416.
\MF{(b) Simulated peak gain at 300 K for different doping densities of the design with 31 \AA~injection barrier.}}
\label{GainAll}
\end{figure}
\begin{figure}
\includegraphics[width=\linewidth]{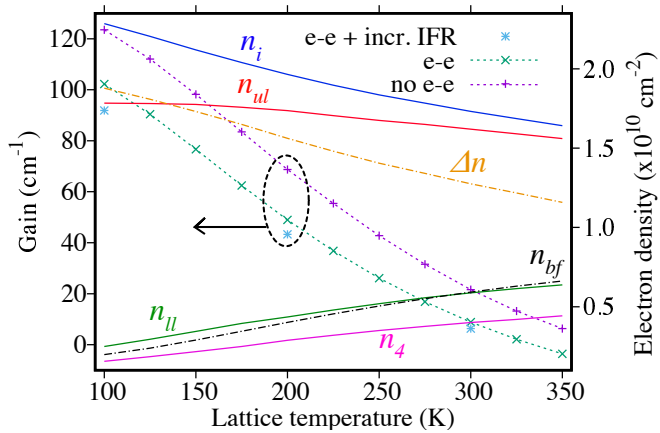}
\caption{(Color online) Simulated gain vs.~lattice temperature for the best design (EV2416). Without electron-electron scattering, the model predicts lasing up to $\sim$280 K. However, including electron-electron scattering reduces the gain significantly, and shifts the maximum operating temperature down by approximately 40 K. The simulated inversion drops much less than gain and essentially follows the change in the lower laser level ($n_{ll}$) by thermal backfilling.}
\label{Tdeg}
\end{figure}

%
In Fig.~\ref{GainAll} (a), the simulated gain is shown for a selected set of layer sequences and doping densities. When doubling the doping density, we see an increase of gain from 50 to 70 cm\inv~at 200 K. Even though the effect is much smaller at 300 K, going from 16 to 20 cm\inv, it still provides significant benefit since gain drops rapidly with temperature (see Fig.~\ref{Tdeg}). In addition, we see that the absorption at higher frequencies gets larger as the population difference between $ll$ and $i$ increases with doping. \MF{For even higher doping densities, as shown in Fig.~\ref{GainAll} (b) for the best design with 31 \AA~injection barrier, the simulated gain is lower at 300 K, and we find an optimal doping density of $\sim$4.5$\cdot10^{10}$ cm\invt. The effect of electron-electron scattering indicates a strong reduction of gain, as well as a shift of the peak gain towards lower doping density.}
Changing the injection barrier width from \MF{the nominal value 34} \AA~to 31 \AA, the $ul$ is more efficiently filled from $i$ and gain increases. For even narrower barrier widths, we see again a decrease in the peak gain (not shown), as thermally activated phonon emission dominates at high temperatures.
The laser barrier width is less relevant, as the change in oscillator strength has a small influence\cite{fathololoumi_terahertz_2012} in this parameter range.

The highest simulated and measured operating temperature was achieved for the structure called EV2416, shown in Fig.~\ref{StatesEV2416}. Here, we see that the phonon extraction has been complemented by a secondary extraction mechanism; tunnelling into state number \MF{$4'$} with subsequent phonon emission landing on the $ul$ \MF{or $i$ state} of the next period, \MF{as indicated by the calculated phonon scattering rates (see Supplementary material, Table 1)}. This resonance is also present in previous 2-well structures\cite{kumar_two-well_2009,scalari_broadband_2010}, where it is detrimental since it has significant overlap with continuum states. This is similar to the situation in Ref.~\onlinecite{albo_temperature-driven_2017}, where the lower laser level was partly depopulated into continuum states rather than a bound state. In contrast, with the higher barriers we employ here, this transition can be safely exploited for increasing inversion\MF{, as we have verified by comparing the energy resolved current densities of our samples with those of Refs.~\onlinecite{kumar_two-well_2009} and \onlinecite{scalari_broadband_2010} (not shown)}.
We found an optimal oscillator strength of $f_\text{osc.} = 0.43$. This value is significantly higher than the previous two-well design of Ref.~\onlinecite{scalari_broadband_2010}, and compares well to the structures with the best THz temperature performance in the literature\cite{fathololoumi_effect_2013}.

For this sample, we also investigate the gain degradation with temperature in detail. The two main inversion degrading mechanisms discussed in literature are thermally activated LO phonon emission and thermal backfilling. The former effect can clearly be seen in the left part of Fig.~\ref{StatesEV2416}, where the states \MF{with in-plane energy $E_k\approx 16$ meV above $ul$, which is precisely one LO phonon energy below $ul'$,} are highly occupied. However, the rate of phonon emission increases by $\sim$ 20\% from 100 K to 300 K, and can only account for a small fraction of the gain degradation. The latter effect can be estimated by comparing the $ll$ population ($n_{ll}$) as a function of temperature, with the one expected from thermal transitions from the highly populated levels $i$ and $ul$. To this end, we show in Fig.~\ref{Tdeg} the occupations of the relevant levels indicated in Fig.~\ref{StatesEV2416}, as well as the expected population ($n_\text{bf}$) of $ll'$, from thermal backfilling. This shows, that thermal backfilling is mainly responsible for the reduction of inversion of our structure. Fig.~\ref{Tdeg} also shows, that the occupation of level 4 roughly follows $n_{ll}$. This indicates that level 4 is depopulating $ll$. This is also evident in the calculated energy resolved current density (see supplementary material), where the effect is much more clear at 300 K than 100 K, indicating a thermally activated process, in agreement with Ref.~\onlinecite{albo_temperature-driven_2017}. 
Simultaneously, the simulations display a
drop in inversion with temperature by 40\% from $T_L = 100$ K to $T_L=300$ K, which can only partially explain the gain drop by 80\% in the same temperature interval. Since the levels $i$ and $ul$ are in resonance, we have defined the inversion as the average population of these levels, minus $n_{ll}$. Over the same temperature range, the FWHM of the main gain peak \emph{decreases}, and thus does not explain the reduction of gain. Possible further sources are the re-absorption by the low-energy tail of the $i\rightarrow ll$ transition, as well as the transition $4\rightarrow 5$. Indeed, the $4\rightarrow 5$ transition energy is only 6 meV below the main one, and the width of level $4$ increases from $\sim$6 meV at 100 K, to $\sim$10 meV at 300 K. In addition, the oscillator strength is very high between levels 4 and 5 due to their spatial overlap\MF{, thus lowering the maximum operating temperature ($T_\text{max}$)}. Similarly, the $i\rightarrow ll$ transition energy is 34 meV and $i$ has a similar width to level $4$. While this transition is further separated from the main one, it is much stronger due to the high occupation of level $i$. Using a simple Fermi's golden rule calculation of the gain, we can \MF{parameterize} the transition broadening independently. By increasing the transition width of all states from 6 meV to 10 meV, using the populations at 300 K, we find a reduction of 75\%, of the peak gain. Our findings thus show that the main part of the gain degradation originates from broadened re-absorption. This could partially be mitigated by moving level $5$ in energy, e.~g.~by the use of higher barriers.

The NEGF simulations \MF{predict} gain as high as 20 cm\inv~at room temperature, which does not agree with the experimental findings discussed below. 
However, the simulations presented above do not include e-e scattering.
In order to check the relevance of this scattering mechanism, we include it within a simplified GW approximation\cite{winge_simple_2016}.
This results in a better thermalization of the electron distribution within the subbands, as seen in the right part of Fig.~\ref{StatesEV2416}. The shorter $ul$ lifetime leads to a reduction of the gain, as seen in Fig.~\ref{Tdeg}. This indicates that neglecting e-e scattering in our simulations leads to an overestimation of the operation temperature. The model includes interface roughness \MF{(IFR)} scattering with a Gaussian correlation function with correlation length $\Lambda = 9$ nm and height $\eta = 0.1$ nm. In order to investigate the sensitivity of the results on these unknown experimental parameters, simulations with $\eta = 0.2$ nm were also carried out. As can be seen in Fig.~\ref{Tdeg}, this further decreases the gain by 5 (2.5) cm\inv~at 200 K (300 K).
\MFtwo{In addition, the simulation temperature
refers to the phonon occupation number. It is known, that the optical
phonons are not in equilibrium\cite{vitiello_non-equilibrium_2012} and thus the effective phonon temperature can be tens of K higher
than the experimental heat-sink temperature even for pulsed operation.}

\begin{figure}
\includegraphics[width=0.95\linewidth]{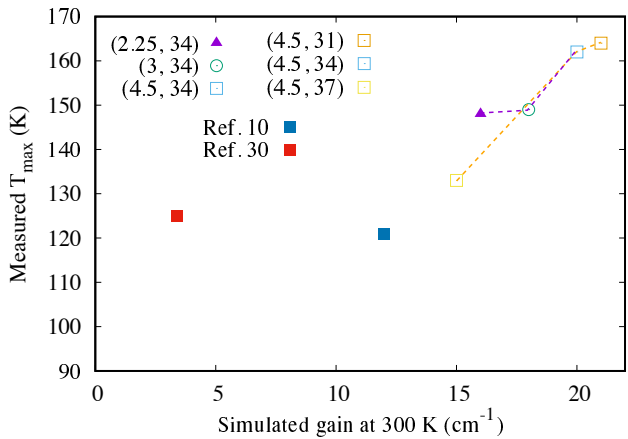}
\caption{(Color online) (a) Measured maximum operating temperature vs.~maximum simulated gain at 300 K, for different structures. The measured devices are processed with wet-etched Au-Au \MF{waveguides}. The symbols indicate doping density (given in units of $10{}^{10}$ cm\invt) and the colour the injection barrier width (given in \AA). The dashed lines are guides for the eye and show the trend of changing doping density and injection barrier width, respectively.}
\label{Trends}
\end{figure}

A selection of structures was characterized experimentally in order to verify the numerical optimization. The designs presented in Fig.~\ref{GainAll} \MF{were} grown by molecular beam epitaxy, and processed into wet-etched Au-Au ridge lasers with varying widths (120-160 $\mu$m) and fixed length of 1 mm. The bottom contact has been etched away before the evaporation of the top metal cladding, in order to reduce the losses due to parasitic absorption. The number of periods were chosen to keep the total thickness of the samples the same (8 $\mu$m).
Fig.~\ref{Trends} shows the maximum operating temperature achieved vs.~the simulated gain at 300 K, which is an indicator of the design optimality. The overall device performance trends from varying barrier thickness and doping density, agree with those of the NEGF simulations. In addition, the current density of the measured samples with varying doping density show the expected trend of increasing current density with doping. The maximum operating temperatures differ widely, from 117 K to 164 K. \MF{Here, we also show the data for the previous 2-well structures, where Ref.~\onlinecite{kumar_two-well_2009} agrees well with the trend from our samples. However, the structure from Ref.~\onlinecite{scalari_broadband_2010} seems to be more temperature sensitive than the other samples. We attribute this to the extraction energy $E_\text{ex}\approx 30$ meV deviating from the optical phonon energy $E_\text{LO}=36.7$ meV for this structure, while both Ref.~\onlinecite{kumar_two-well_2009} and our designs have $E_\text{ex}\approx E_\text{LO}$.}

\begin{figure}
\includegraphics[width=\linewidth]{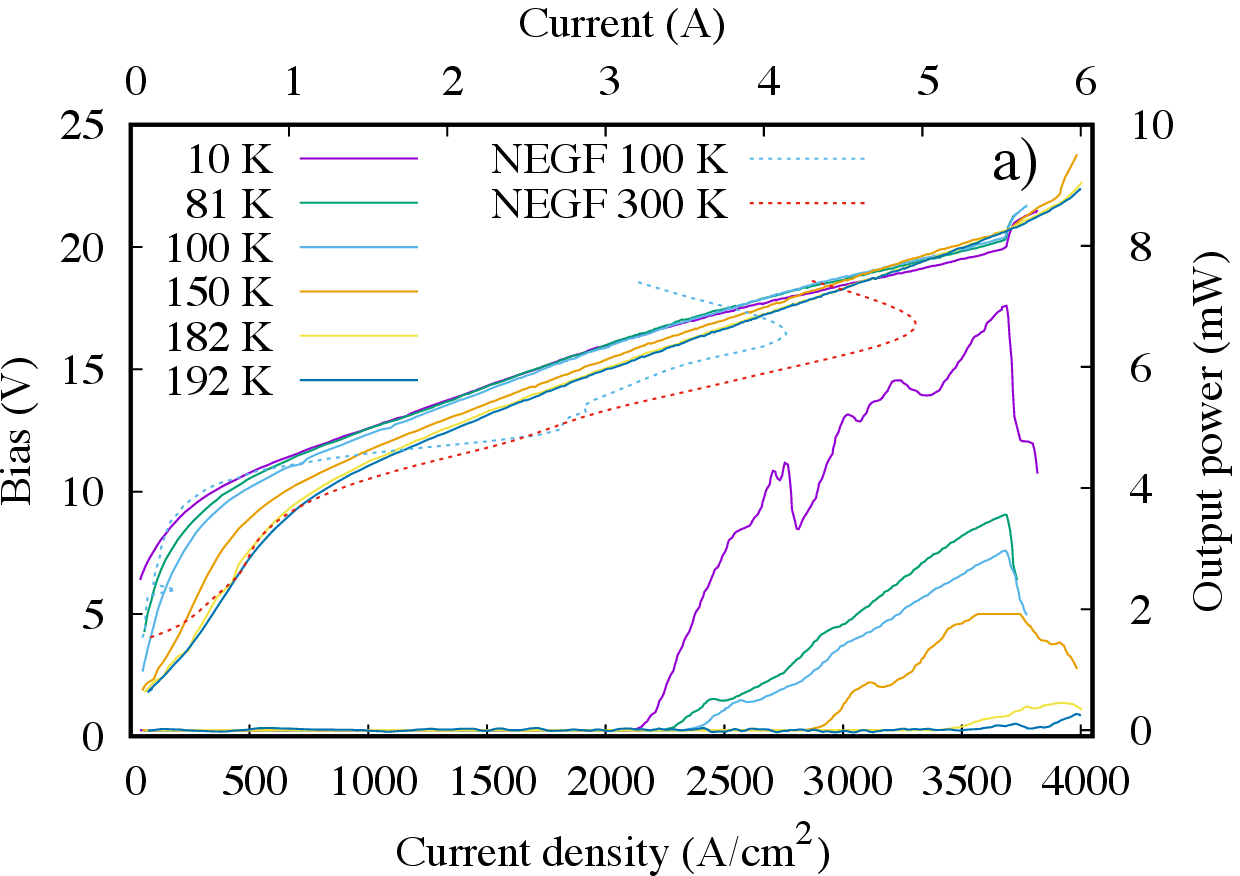}\\
\vspace*{0.2cm}
\includegraphics[width=\linewidth]{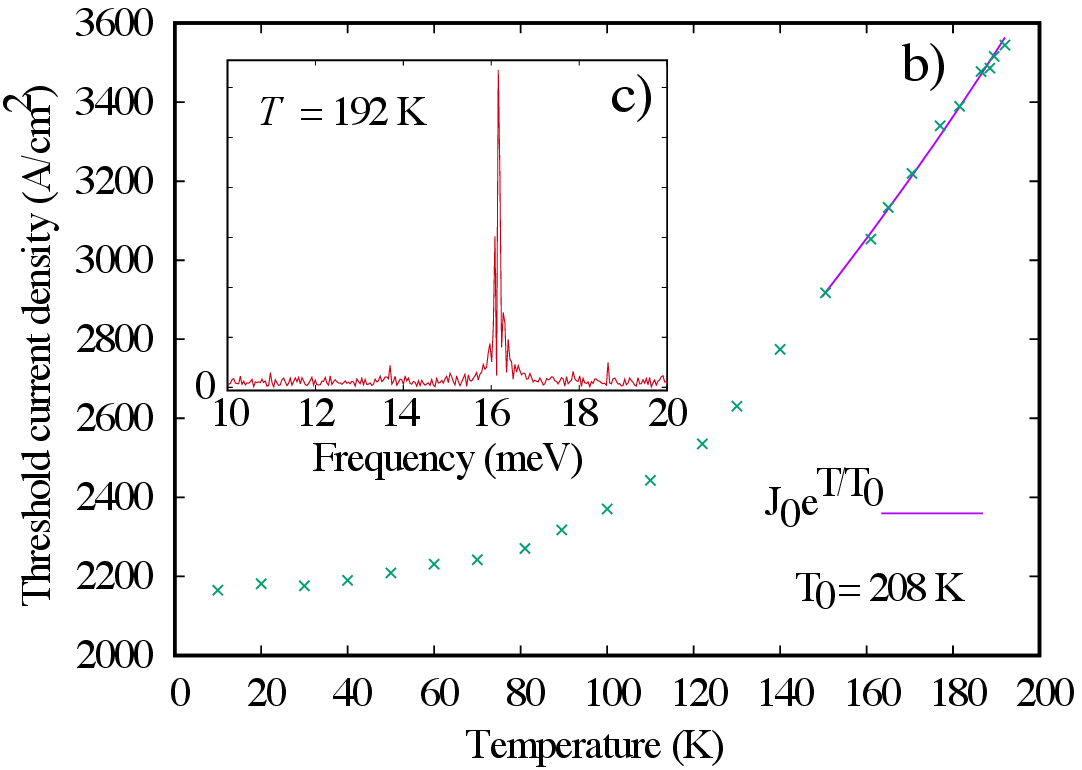}
\caption{(Color online) (a) Pulsed LIV at different temperatures\MF{, for a 1 mm long, dry-etched Cu-Cu process of} the best device labelled EV2416. Dashed lines show NEGF simulations, shifted by 3.8 V, at 100 K and 300 K. (b) Threshold current vs. temperature for the best design. The fit gives a $T_0$ of 208 K. (c) Laser spectrum for a 2 mm long device near the maximum operating temperature.}
\label{EV2416exp}
\end{figure}

For 1 mm long Au-Au waveguides, we have simulated the waveguide and mirror losses from a time-domain spectroscopy (TDS) measurement of the transmission of a sample including the top contact of our laser and a 50 nm Au layer. This calculation gives waveguide losses of 30 cm\inv. The mirror losses are calculated to be 4 cm\inv, and thus we estimate a threshold gain of approximately 35 cm\inv~at 200 K. The NEGF simulations predict a lowest $T_\text{max}$ of 218 K, with e-e and increased interface roughness \MF{included}. This is 54 K above the measured $T_\text{max}$. However, it is worth to note that the simulations do not include effects such as Joule heating and non-equilibrium phonons. In addition we did not consider the absorption of the tail of the TO phonon resonance.
Together with gain optimisation, waveguide losses \MF{also} need to be minimized.
To this end, the best sample (EV2416) was also processed with a dry etched Cu-Cu waveguide, which is expected to have lower losses\cite{belkin_terahertz_2008}. The characterisation of LIV is shown in Fig.~\ref{EV2416exp} (a). The best device (1 mm long, 140 $\mu$m wide) operated up to a temperature of 192 K, and showed a high $T_0$=208 K, as shown in Fig.~\ref{EV2416exp} (b). We also show in Fig.~\ref{EV2416exp} (a) the simulated current density in the NEGF model, which has been shifted by an assumed potential drop of 3.8 V, due to a Schottky contact. The laser spectrum measured at 192 K is shown in Fig.~\ref{EV2416exp} (c) and the lasing frequency agrees with the simulated gain spectrum. However the maximum current density is underestimated in the NEGF model, even at high temperature where photo-driven current is negligible. Including e-e scattering in the simulations, we find a maximum current density of 3.3 kA/cm\inv at 200 K. While this agrees better with the experiment, it cannot account for the high experimental current density at 190 K. This indicates that continuum leakage is still present at temperatures where highly excited states become thermally occupied. Together with the high simulated gain and the high $T_0$, this suggests that the excellent laser performance of the presented design can be further improved.

In conclusion, we have optimized 2-well QCLs using a combination of complex numerical simulations, and experimental measurements. We find an optimal structure featuring both phonon and resonant tunnelling extraction and injection. The agreement between the experimental and simulated trends highlights the efficacy of our model for optimization of QCL structures. Together with a Cu-Cu waveguide to reduce optical losses, we have significantly improved the operation temperature of 2-well THz QCLs, close to the overall record temperature. 
We see potential to further improve the temperature performance of THz QCLs; the doping density, material parameters (such as barrier height), as well as optical losses can be further optimized. The main gain degradation mechanism at high temperature was found to be temperature broadening of re-absorption transitions, while thermal backfilling is responsible for the reduction of inversion. The effect of electron-electron scattering was found to be significant, reducing the maximum operating temperature by $\sim$ 40 K. Including this scattering mechanisms in more detail, may therefore be helpful for further optimization.

\section*{Supplementary material}
In order to clearly show the presence of the tunnelling extraction channel, we present the energetically and spatially resolved current densities for varying temperature\MF{, as well as the relevant calculated LO phonon scattering rates.}

This project has received funding from the European Research Council (ERC) under the project MUSiC. A.~W.~acknowledges the Swedish Research Council (VR) for financial support. The simulations were partially performed on resources provided by the
Swedish National Infrastructure for Computing (SNIC) at LUNARC.

%


\end{document}